# Scalable transparent conductive thin films with electronically passive interfaces for direct chemical vapor deposition of 2D materials

Theresa Grünleitner,[a,‡] *Alex Henning,*[a,‡,*] *Michele Bissolo,*[a] *Armin Kleibert,*[b] *Carlos A.F. Vaz,*[b] *Andreas V. Stier,*[a] *Jonathan J. Finley,*[a] *and Ian D. Sharp*[a,*]

[a] Walter Schottky Institute and Physics Department, Technical University of Munich, 85748 Garching, Germany.

[b] Swiss Light Source, Paul Scherrer Institute, CH-5232 Villigen PSI, Switzerland.

ABSTRACT: We present a novel transparent conductive support structure for two-dimensional (2D) materials that provides an electronically passive 2D/3D interface while also enabling facile interfacial charge transport. This structure, which comprises an evaporated nanocrystalline carbon (nc-C) film beneath an atomic layer deposited alumina (ALD $AlO_x$) layer, is thermally stable and allows direct chemical vapor deposition (CVD) of 2D materials onto the surface. When the nc-C/$AlO_x$ is deposited onto a 270 nm $SiO_2$ layer on Si, strong optical contrast for monolayer flakes is retained. Raman spectroscopy reveals good crystal quality for $MoS_2$ and we observe a ten-fold photoluminescence intensity enhancement compared to flakes on conventional $Si/SiO_2$. Tunneling across the ultrathin $AlO_x$ enables interfacial charge injection, which we demonstrate by artifact-free scanning electron microscopy and photoemission electron microscopy. Thus, this combination of scalable fabrication and electronic conductivity across a weakly interacting 2D/3D interface opens up new application and characterization opportunities for 2D materials.

TEXT: Two-dimensional (2D) van der Waals (vdW) materials offer significant promise for application in advanced optoelectronics, quantum technologies, and catalysis.[1–3] While the extreme sensitivity of 2D materials to the surrounding environment opens unique application prospects,[4,5] device functionality can be limited by the availability of substrates possessing a specific combination of optical, electronic, and chemical properties. Currently, $SiO_2$ films on Si serve as the substrates of choice for both chemical vapor deposited and mechanically transferred 2D materials.[6–9] One major benefit of these $Si/SiO_2$ substrates is that strong optical contrast between mono- and few-layer 2D materials can be achieved by controlling the thickness of the $SiO_2$ dielectric film, thereby enabling flakes to be spatially located and the number of layers to be readily determined using non-destructive optical microscopy.[10–13] While these substrates offer advantages for numerous 2D device applications and enable back-gating of 2D transistors,[8,14] they also suffer from small gate capacitances defined by the large oxide thickness required to achieve optimal optical contrast. Furthermore, they are poorly suited for realizing electrically driven optical devices and are incompatible with several common characterization techniques, such as X-ray photoelectron spectroscopy (XPS), scanning electron microscopy (SEM), and Kelvin probe force microscopy (KPFM), which rely on fast carrier discharging. These critical gaps can be overcome by the development of a new interface structure that provides optical transparency, tunable gate impedance, and broad compatibility with different substrates and processes, including for direct chemical vapor deposition (CVD) of 2D materials on the surface. Since such support structures must also be chemically inert and ultra-smooth, simultaneously fulfilling all of these requirements represents a significant materials challenge.

Although transparent conductive films (TCFs), such as indium tin oxide (ITO) and fluorine-doped tin oxide (FTO),[15] are key components in a broad range of optoelectronic devices,[15–18] they are susceptible to high temperature instabilities, rendering them unsuitable as substrates for 2D material growth. In addition, TCFs typically have relatively high surface roughness (~1 nm),[16] whereas ultra-smooth surface morphologies and potential landscapes are required for 2D semiconductors to achieve high carrier mobilities, homogeneous carrier densities and reproducible (opto)electronic characteristics.[19] To overcome these limitations, graphene-based TCFs were recently investigated for integration into field effect transistors (FETs) and optical modulators.[20] Although these systems can exploit the high electrical conductivity, mechanical flexibility, and optical transparency of graphene,[20] single graphene layers have limited sheet carrier concentrations that can be easily screened, limiting their application as gate electrodes. Furthermore, fabrication of large-area graphene-based electronic devices is complex, often requiring transfer processes that can introduce contaminants and structural defects.[21,22]

Here, we present a novel TCF structure based on conductive and transparent nanocrystalline carbon (nc-C) films that overcome the limitations of existing substrates. Although studied for many decades, nc-C remains largely underexplored as a TCF for 2D materials applications. We show that nc-C substrates can be produced on a large scale with tunable sheet carrier concentrations. We conformally coat the nc-C layer with an amorphous aluminum oxide ($AlO_x$) layer, having a controlled thickness (down to 1 nm and even below[23,24]) using atomic layer deposition (ALD). While the conductive ($\rho < 0.01$ $\Omega$ cm) and ultra-smooth (~100 - 150 pm) carbon coating serves as the transparent electrode, the conformal alumina layer acts as a high-$k$ dielectric spacer with tunable impedance. Of particular relevance to 2D materials, this structure is thermally and chemically stable under typical growth conditions for CVD of transition metal dichalcogenides, with the alumina layer facilitating the nucleation and growth of single layer $MoS_2$ flakes. Importantly, when the nc-C/$AlO_x$ film is deposited on a Si substrate coated with 270 nm $SiO_2$, the high optical contrast for discerning single and multi-layer flakes is retained, thereby enabling subsequent analysis and processing. To demonstrate the utility of this structure, we show that 2D flakes can be directly characterized by electron microscopy and spectroscopy techniques, including scanning electron microscopy (SEM) and synchrotron-based X-ray photoemission electron microscopy (XPEEM), without charging effects that would otherwise prevent such measurements on conventional Si/$SiO_2$ supports. This is possible because the $AlO_x$ thickness is within the length scale of tunneling, thereby enabling rapid discharging of the surface layers. Raman spectroscopy shows good crystal quality of $MoS_2$ and photoluminescence (PL) measurements reveal that the TCF support yields significantly enhanced PL emission intensity from the $MoS_2$ compared to $SiO_2$ substrates, which is attributed to effective electronic shielding of charged defects within $SiO_2$. Beyond the application of the nc-C/$AlO_x$ TCF for 2D materials, the presented scheme provides a highly versatile and flexible route to the uniform and scalable deposition of TCFs on other substrates of relevance to thin film optical and electronics applications.

Figure 1a illustrates the smooth, conductive, and optically transparent nc-C/$AlO_x$ structure (hereafter also referred to as the TCF structure), which was deposited onto a Si substrate covered by a 270 nm thick layer of thermal $SiO_2$. We note that the Si/$SiO_2$ substrate was selected here as a model support that is commonly used to provide excellent optical contrast in 2D materials research, although alternative substrates could also be utilized. The as-deposited structure was probed using atomic force microscopy (AFM) and found to have a low surface roughness of 211 pm.

Following fabrication of the $Si/SiO_2/nc\text{-}C/AlO_x$ structure, $MoS_2$ was grown on the surface *via* atmospheric pressure chemical vapor deposition (CVD) at 850°C. Importantly, we find that the complete structure is thermally and chemically stable, withstanding the sulfidizing environment at elevated temperatures during CVD (see below), with the $AlO_x$ layer promoting the nucleation and growth of primarily single layer triangular $MoS_2$ flakes across its surface, as indicated by the optical micrograph shown in Figure 1b. In addition to confirming the successful growth of $MoS_2$, this micrograph highlights that the strong optical contrast between 2D materials and the substrate is retained, despite the presence of the nc-C/$AlO_x$ TCF structure. Indeed, quantification of this contrast $C = \left|\frac{G_{\text{sub}} - G_{\text{MoS2}}}{G_{\text{Sub}}}\right|$,[25] where $G_{\text{sub}}$ and $G_{\text{MoS2}}$ are the substrate and $MoS_2$ intensities of the green channel of the RGB mix, respectively, yields values of $C_{\text{TCF}} = 0.28$ for $Si/SiO_2/nc\text{-}C/AlO_x$ (Figure 1b) and $C_{\text{Si/SiO2}} = 0.26$ for standard $Si/SiO_2$ substrates (Figure S1). The retention of strong optical contrast between the 2D sheets and the underlying substrate is an important feature of the nc-C/$AlO_x$ TCF that arises from its optically transparency (Figure S2) and ultrathin structure.

Complementary AFM measurements (Figure 1c) confirm the dominant presence of monolayer, crystalline $MoS_2$ flakes, along with a smaller fraction of bi- and tri-layer regions. Furthermore, AFM reveals that the smooth surface morphology of the underlying TCF structure is preserved after CVD growth, with a roughness of 160 pm that is likely determined by the height variations of the chemomechanically polished $Si/SiO_2$ substrate (~160 pm). We emphasize that a smooth and chemically inert substrate surface is necessary to preserve the intrinsic properties of few-layer and monolayer 2D materials because it reduces microscopic strain, while also facilitating the release of strain without distorting the 2D lattice during cooling from the CVD growth temperature.[26]

For comparison purposes, a 3 nm thin gold coating, deposited by electron-beam evaporation onto a Si/SiO$_2$ substrate, provides a similar conductivity ($\rho$ = 0.01 $\Omega$ cm) to the nc-C/AlO$_x$ TCF structure, but has a much higher roughness (~1.1 nm) and absorbs light more strongly, thereby reducing the optical contrast of monolayer MoS$_2$ to 0.17. This comparison highlights the advantages of using nc-C, rather than a metal, as the conductive layer in such a structure.

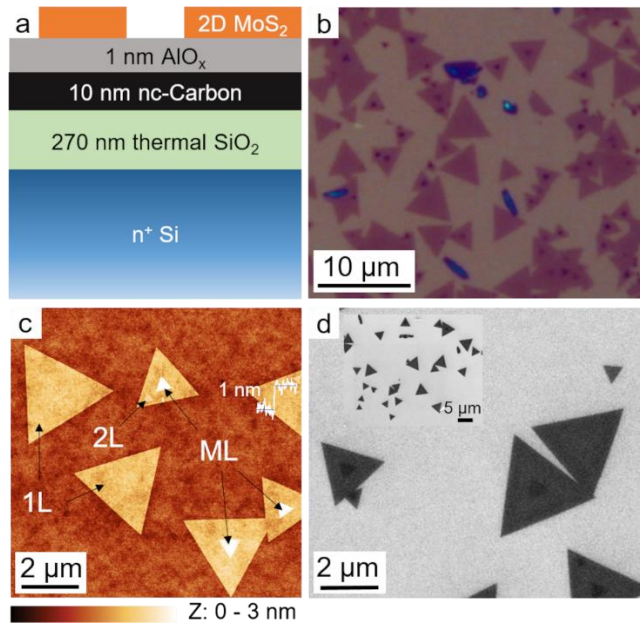

**Figure 1.** (a) MoS$_2$ (orange) on n$^+$ Si/thermal SiO$_2$ substrate coated with a transparent conductive film (TCF) composed of nanocrystalline carbon (nc-C) capped with a 1 nm thin aluminum oxide (AlO$_x$) layer. The full layer stack is thus given by Si/SiO$_2$/nc-C/AlO$_x$/MoS$_2$. (b) Optical microscopy image of 2D MoS$_2$ on the structure depicted in (a), demonstrating that high optical contrast characteristic of the 270 nm thick SiO$_2$ and facile layer thickness determination is retained, despite the presence of the conductive nc-C film. (c) AFM image of MoS$_2$ flakes on the structure, indicating the ultra-smooth and closed nature of the surface following growth of MoS$_2$ via CVD.

The inset shows the corresponding height profile. (d) SEM image of MoS$_2$ on the Si/SiO$_2$/nc-C/AlO$_x$ structure, which enables direct high-resolution imaging without electrical charging effects. The zoomed-out image in the inset further demonstrates the lack of charging-induced artifacts during SEM imaging.

In addition to facilitating CVD growth of MoS$_2$, the 1 nm thin AlO$_x$ coating enables charge carriers to tunnel from MoS$_2$ to the conductive carbon film. As a consequence, electron microscopy and spectroscopy methods can be readily applied for the characterization of the 2D material. Figure 1d shows a scanning electron microscopy (SEM) image of individual monolayer MoS$_2$ flakes on top of the Si/SiO$_2$/nc-C/AlO$_x$ structure. No artifacts arising from charging of the 2D material are observed, including at lower magnification (see inset) after initial high-resolution imaging. The resulting SEM images are characterized by sharp features and reveal bilayer nucleation sites within individual monolayer flakes (Figure 1d) that were not always discernable using optical microscopy. In contrast, SEM of MoS$_2$ on a standard Si/SiO$_2$ substrate displays significant charging artifacts, which greatly reduce image quality and resolution (Figure S3).

The utility of the TCF structure for facilitating nanoscale characterization of 2D materials is further highlighted by X-ray photoemission electron microscopy (XPEEM), which is a powerful spectro-microscopic method for nanoscale composition and chemical analysis of surfaces, but requires efficient discharging of the studied material. The Si/SiO$_2$/nc-C/AlO$_x$ substrate presented here fulfils this requirement and enables high resolution XPEEM analysis of the local chemical composition of CVD-grown MoS$_2$ flakes, as shown in Figures 2a and b, in which each pixel represents the integrated area of a single Mo 3d 5/2 (Figure 2a) or Al 2p (Figure 2b) peak after background subtraction. The large elemental contrast and sharp edges between the MoS$_2$ triangles (green) and substrate (dark blue) in Figure 2a confirm the lack of electrostatic charging during XPEEM of MoS₂ grown on TCF-coated Si/SiO₂. The smaller sub-µm sized high-intensity features (yellow) represent bilayer nucleation spots on monolayer MoS₂ flakes, also observed in AFM and SEM. The Al 2p elemental contrast map (Figure 2b) demonstrates the homogeneity of the AlO$_x$ coating of the substrate (green).

The chemical state and purity of CVD MoS₂ on the TCF is further shown by a typical Mo 3d and S 2s XPS core level spectrum obtained from a single flake (Figure 2c). The spectrum is dominated by the spin-orbit split doublet at 230.5 eV and 233.7 eV, corresponding to pristine Mo-S bonding within the material, along with the S 2s signal at 227.8 eV. The weak doublet having components at 233.8 eV and 237.0 eV is attributed to Mo-O and has a concentration of < 3.5%, which is well below the oxygen concentrations of intentionally doped MoS₂ films reported by Wei et al.[27] These findings reveal that the 2D MoS₂ is of good quality with minimal oxygen incorporation, doping, or contamination from other chemical species.

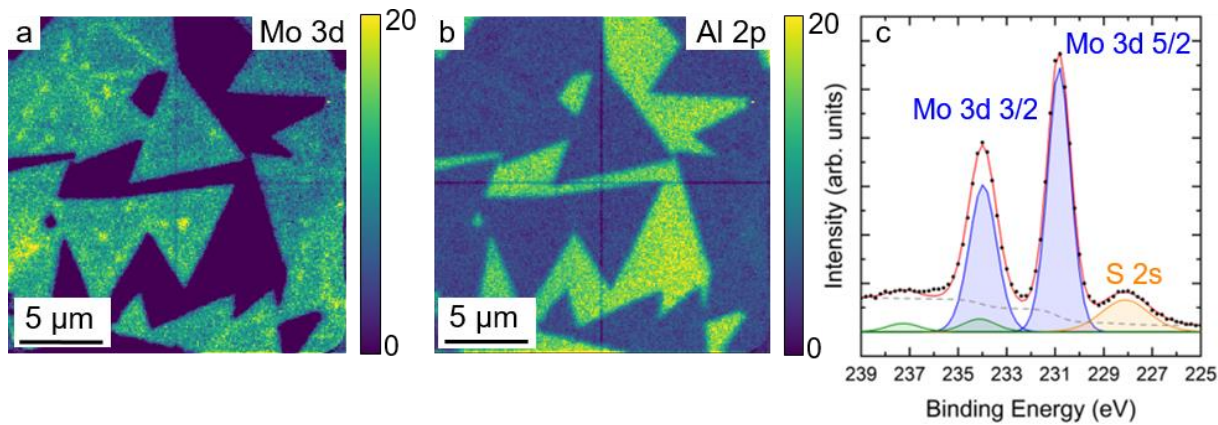

**Figure 2.** X-ray photoelectron maps of MoS₂ collected from the (a) Mo 3d and (b) Al 2p core-level regions. Acquisition of these images with negligible charging was enabled by the Si/SiO₂/nc-C/AlO$_x$ support, which allows rapid discharging of the surface layers via the conductive nc-C layer. (c) A typical Mo 3d–S 2s core level spectrum extracted from a sequence of maps acquired with the photon energy ranging from 238.7 – 224.7 eV, simultaneously indicating the high

chemical quality of the MoS$_2$ layer grown by CVD and the possibility for high spatial and energy resolution photoemission from supported flakes. Fitted components are presented in blue (Mo3d-S), green (Mo3d-O), and orange (S2s-Mo), respectively.

To further analyze the structural and optoelectronic quality of the CVD MoS$_2$ on the TCF, we compare Raman and photoluminescence spectra obtained from individual MoS$_2$ nanosheets on Si/SiO$_2$/nc-C/AlO$_x$ substrates to those on standard Si/SiO$_2$ surfaces. Raman measurements recorded on both substrates (Figure 3a) show that the A′$_1$ and E′ Raman modes of the flakes are separated by $\Delta_{TCF}$ = 20.94 ± 0.14 cm$^{-1}$ ($\Delta_{Si/SiO_2}$ = 20.60 ± 0.10 cm$^{-1}$), which are characteristic of monolayer MoS$_2$. The narrow linewidths of the two Raman spectral features, $FWHM_{TCF,A'_1}$ = 6.16 ± 0.40 cm$^{-1}$ and $FWHM_{TCF,E'}$ = 5.13 ± 0.30 cm$^{-1}$ ($FWHM_{Si/SiO_2,A'_1}$ = 5.11 ± 0.25 cm$^{-1}$ and $FWHM_{Si/SiO_2,E'}$ = 4.32 ± 0.08 cm$^{-1}$), are fully consistent with previous reports of exfoliated and CVD-grown monolayer MoS$_2$.[28,29] These observations reaffirm the high crystal quality of MoS$_2$ grown on the TCF structure. Since comparison of the peak positions reveals no shift of the Raman modes within the error ($\Delta_{E'}$ = 0.03 cm$^{-1}$, $\Delta_{A'_1}$ = 0.31 cm$^{-1}$) for MoS$_2$ on the TCF-coated substrate relative to MoS$_2$ on bare Si/SiO$_2$, we conclude that an equivalent quality of MoS$_2$ flakes is achieved on both substrates. Finally, we note that the spectral features at approximately 380 cm$^{-1}$ and 409 cm$^{-1}$ that appear on the left-hand (right-hand) side of the E′ (A$_1$′) mode are present in both spectra with similar intensity (Figure 3a). These features are reportedly related to disorder-induced Raman modes at the M point, indicating a similar degree of structural disorder within flakes on both substrates.[30,31]

We continue our analysis of substrate-related optical properties of CVD MoS$_2$ with PL spectroscopy, which allowed us to probe substrate-induced effects with a higher sensitivity than Raman spectroscopy.[32] Figure 3b compares PL spectra recorded from the TCF and SiO$_2$ substrates when subject to comparable excitation conditions. Remarkably, the PL intensity for monolayer MoS$_2$ on the TCF surface increases by more than one order of magnitude compared to the PL intensity of MoS$_2$ on the regular Si/SiO$_2$ substrate. To determine the origin of this PL enhancement, we first investigated the influence of the substrate terminal surface by comparing the PL from MoS$_2$ on Si/SiO$_2$ to that on Si/SiO$_2$/AlO$_x$ (Figure S4). Even in the absence of the nc-C layer, we observed a significant (~5×) increase of the PL intensity on AlO$_x$, which suggests reduced substrate dangling bond defects and covalent interactions.[33] With the addition of the nc-C layer to form Si/SiO$_2$/nc-C/AlO$_x$/MoS$_2$, we observed an even larger enhancement of the PL intensity, which is consistent with additional screening of charged defect states that are ubiquitous in SiO$_2$ by the conductive nc-C layer.[34] Together, these effects yield a ~10× PL enhancement of MoS$_2$ on the TCF compared to our reference substrate SiO$_2$.

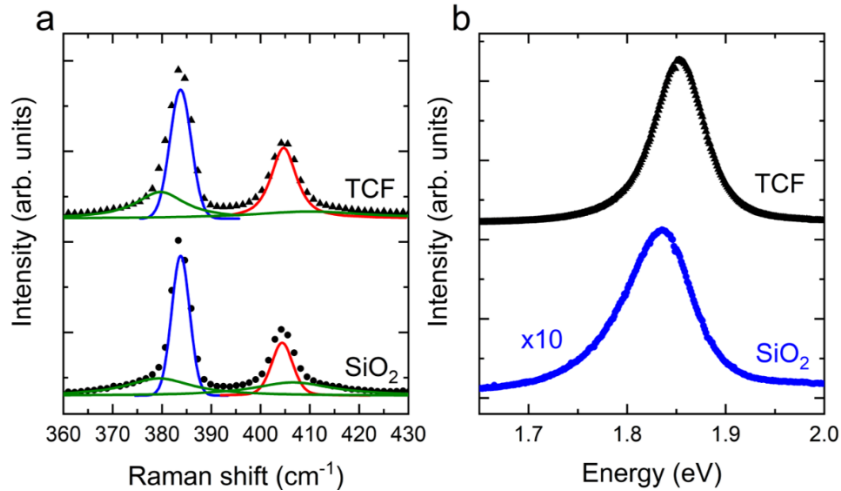

**Figure 3.** (a) Raman and (b) PL spectra of representative, individual MoS$_2$ flakes on the nc-C/AlO$_x$ TCF structure (triangles/black, respectively) and on SiO$_2$ (circles/blue, respectively) acquired with identical excitation conditions (532 nm, 0.6 MW/cm2 CW power) at room temperature. The normalized Raman data in (a) show the fitted E´ mode in blue and the A1´ mode in red. The green features are related to disorder-induced Raman modes. The vertical lines serve as a guide to the eye. To aid comparison, the PL spectrum of MoS$_2$ on Si/SiO$_2$ in (b) has been magnified by a factor of 10×. The spectra are shifted vertically for better comparison.

Having demonstrated the applicability of this structure for the growth and characterization of high quality MoS$_2$, we now turn to the analysis of the basic properties of the nc-C/AlO$_x$ structure, noting that it offers a combination of characteristics that can be broadly useful, potentially reaching beyond 2D material applications. The first fabrication step comprises room temperature deposition of a conformal 10 nm thick amorphous carbon layer, which is electrically resistive ($\rho > 10$ Ω cm) and smooth (rms roughness 173 pm) in its as-deposited state, as summarized in Table 1. As discussed below, the resistivity of this layer is significantly reduced *via* thermal annealing, including during CVD growth of MoS$_2$.

Although ALD AlO$_x$ has been previously demonstrated for growth on top of van der Waals materials,[35,36] this can be challenging since a lack of binding sites could inhibit growth on the evaporated carbon. Thus, we used *in situ* spectroscopic ellipsometry to track the alumina thickness during ALD on the carbon film with Å-level precision (Figure S5). The lack of reactive sites on the carbon surface can explain the observed delayed film nucleation (Figure S5), requiring chemical activation during the first several ALD cycles. However, following this induction period, deposition of a nanometer-thin conformal AlO$_x$ coating on the as-deposited amorphous carbon layer was achieved.

As summarized in Table 1, growth of the ALD layer does not have a significant impact on the high resistivity of the as-deposited carbon layer, while the roughness is slightly increased to 211 pm. However, we find that annealing at 800 °C in Ar atmosphere transforms the material to a smooth nanocrystalline layer with significantly lower resistivity of $1.2\times10^{-2}$ Ω cm (Table 1). As shown in Figure S2, the decreased resistivity is accompanied by a moderate increase of optical absorption, although the material remains optically transparent over a broad spectral range.

To better understand the effect of thermal processing on the structure and resistivity of the carbon film, we performed Raman spectroscopy at different stages in the fabrication of the nc-C/AlO$_x$ TCF. Figure 4 compares the two dominant Raman spectral features of the carbon film, the G peak associated with sp$^2$ carbon stretching modes and the D peak associated with a breathing mode of the ring-shaped sp$^2$ bonds present in disordered graphite.[37,38] From analysis of the Raman peak positions and the intensity ratio of the D peak relative to the G peak, I(D)/I(G), one can assess the degree of disorder in the carbon film and estimate the cluster diameter or cluster correlation length, $L_a$.[39,40] As illustrated in Figures 4a,4b the peak positions as well as the intensity ratio, I(D)/I(G), remained nearly constant after AlO$_x$ growth at 200 °C, suggesting that the carbon structure was unaffected by the ALD process. In contrast, after annealing in an Ar atmosphere at 800 °C for 5 minutes, we observe a strong blueshift of the G peak and an increase of the I(D)/I(G) ratio (Figure 4c & Table 1). Together, these changes indicate formation of graphite nanocrystals *via* transformation of sp$^3$ bonds to sp$^2$ bonds,[37] as well as clustering of sp$^2$ rich regions.[41] Given prior Raman studies of carbon thin films,[38] we conclude that the carbon layer is predominantly composed of nanocrystalline graphite (sp$^2$ bonds with $L_a$ ~40 Å) with less than 20% sp$^3$-bonded atoms. This dominant content of graphitic sp$^2$ carbon clusters is consistent with the significantly increased conductivities of annealed films[42,43] (Table 1), as well as the partially reduced optical transparency (Figure S2). Importantly, performing CVD growth of MoS$_2$ yields nearly identical Raman spectra as annealing in Ar (Figure 4d & Table 1 and XPEEM), indicating that the conversion of the films to graphitic nc-C occurs similarly under the sulfudizing growth conditions.

As a consequence, no pre-annealing of the substrate was required to achieve the conductive carbon layer during fabrication of the complete Si/SiO$_2$/nc-C/AlO$_x$ structure.

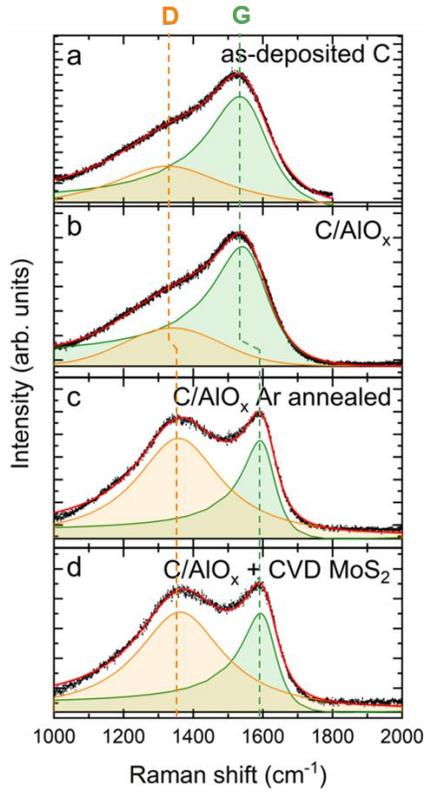

**Figure 4.** Raman measurements of (a) as-deposited C, (b) C/AlO$_x$, (c) C/AlO$_x$ annealed in Ar atmosphere, and (d) C/AlO$_x$ with CVD-grown MoS$_2$. The solid red lines represent the cumulated fitted peaks. The G peak is depicted in green, the defect peak D in orange. The vertical dotted lines are a guide to the eye. The TCFs were deposited on Si/SiO$_2$ substrates.

**Table 1**: Extracted values from Raman measurements, resistivity, and roughness.

|  |  | as-deposited C | C/AlO$_x$ | C/AlO$_x$ Ar annealed | C/AlO$_x$ +CVD MoS$_2$ |
|---|---|---|---|---|---|
| G peak [a] | peak position / cm$^{-1}$ | 1545.58 | 1549.00 | 1599.31 | 1600.84 |
|  | FWHM / cm$^{-1}$ | 109.17 | 96.74 | 57.12 | 60.31 |
| D peak [a] | peak position / cm$^{-1}$ | 1326.88 | 1335.27 | 1358.10 | 1362.41 |
| | FWHM / cm$^{-1}$ | 415.32 | 328.56 | 297.13 | 301.80 |
| | I(D)/I(G)[a] | 0.41 | 0.37 | 1.09 | 1.08 |
| | Resistivity / Ω cm | 1.5·10$^6$ | 4.7·10$^1$ | 1.2·10$^{-2}$ | - |
| | rms roughness /pm | 173 | 211 | 156 | 160 |

[a] extracted from Raman measurements

In summary, we have developed a scalable support structure for 2D materials consisting of nanocrystalline carbon coated with an amorphous and conformal ALD-grown AlO$_x$ layer. The structure is smooth, conductive, and optically transparent. The ultrathin conformal AlO$_x$ coating facilitates direct CVD growth of 2D MoS$_2$ on the surface, while providing an electronically passive 2D/3D interface that also enables facile interfacial charge transport. While the nc-C/AlO$_x$ structure can be applied to various passive or functional substrates, its integration with conventional Si/SiO$_2$ supports allows the strong optical contrast between single- and multi-layer 2D materials to be retained, while also yielding a strong photoluminescence enhancement. This latter effect is a consequence of screening of charged defect states in the SiO$_2$ by the conductive nc-C layer. Since ALD allows fabrication of conformal AlO$_x$ high-*k* dielectric spacers of arbitrary thickness down to 1 nm, rapid interfacial charge tunneling across the interface is possible. This feature enables physical, electronic, and chemical characterization using electron imaging and spectroscopy techniques that are otherwise complicated on oxidic supports. As two examples of this, SEM imaging and XPEEM spectromicroscopy are applied to CVD-grown MoS$_2$ on Si/SiO$_2$/nc-C/AlO$_x$ supports that simultaneously provide high optical contrast for single layer flakes. In a wider context, the tunability of ALD dielectric coatings provides the opportunity to reliably control interfacial impedance over a wide range while retaining electronically passive and weakly interacting interfaces. Furthermore, the deposition process for scalable fabrication of nc-C/AlO$_x$ heterostructures is expected to be broadly compatible with diverse substrates and suitable for applications in numerous optoelectronic applications, extending beyond 2D materials.[44,45]

ASSOCIATED CONTENT

**Supporting Information**. The following files are available free of charge. Experimental details, optical microscopy images of MoS$_2$ on different substrates, resistivity and transmission spectra of novel TCF on sapphire, SEM, PL comparison of MoS$_2$ on different substrates, and *in situ* spectroscopic ellipsometry of ALD AlO$_x$ deposition. (Word)

AUTHOR INFORMATION

**Corresponding Authors**

*Ian D. Sharp, Walter Schottky Institute and Physics Department, Technical University of Munich, 85748 Garching, Germany; Email: sharp@wsi.tum.de.

*Alex Henning, Walter Schottky Institute and Physics Department, Technical University of Munich, 85748 Garching, Germany; Email: Alex.Henning@wsi.tum.de

**Author Contributions**

T.G., A.H., and I.D.S designed the study. T.G and A.H. prepared the samples. T.G. performed and analyzed optical microscopy, SEM, Raman, and photoluminescence experiments. A.H. and T.G. performed AFM measurements. T.G., A.H., M.B., A.K., and C.A.F.V. performed and T.G. and M.B. analyzed XPEEM measurements. A.H. and M.B. performed measurements for electrical characterization. A.V.S took part in the discussion of the results. T.G., A.H., and I.D.S. wrote the manuscript. I.D.S., A.H., and J.J.F. supervised the study. All authors reviewed and commented on the manuscript.

‡T.G. and A.H. contributed equally to this work.

**Notes**

The authors declare no competing financial interest.

ACKNOWLEDGMENTS

This work was supported by the DFG through the TUM International Graduate School of Science and Engineering (IGSSE), project FEPChem2D (16.01), and by the Deutsche Forschungsgemeinschaft (DFG, German Research Foundation) under Germany´s Excellence Strategy – EXC 2089/1 – 390776260. AH acknowledges funding from the European Union's Horizon 2020 research and innovation programme under the Marie Skłodowska-Curie grant agreement No 841556. We thank J. Primbs for help with the conductivity measurements, M. Zengerle for her help with sample preparation, and S. Wörle for his help with SEM measurements. Part of this work was performed at the Surface/Interface: Microscopy (SIM) beamline of the Swiss Light Source, Paul Scherrer Institut, Villigen, Switzerland.

REFERENCES

1. Mak, K. F. & Shan, J. Photonics and optoelectronics of 2D semiconductor transition metal dichalcogenides. *Nature Photon* **10**, 216–226 (2016).

2. Klein, J. *et al.* Engineering the Luminescence and Generation of Individual Defect Emitters in Atomically Thin $MoS_2$. *ACS Photonics* **8**, 669–677 (2021).

3.	Parzinger, E. *et al.* Hydrogen evolution activity of individual mono-, bi-, and few-layer MoS$_2$ towards photocatalysis. *Applied Materials Today* **8**, 132–140 (2017).

4.	Florian, M. *et al.* The Dielectric Impact of Layer Distances on Exciton and Trion Binding Energies in van der Waals Heterostructures. *Nano Lett.* **18**, 2725–2732 (2018).

5.	Stier, A. V., Wilson, N. P., Clark, G., Xu, X. & Crooker, S. A. Probing the Influence of Dielectric Environment on Excitons in Monolayer WSe$_2$: Insight from High Magnetic Fields. *Nano Lett.* **16**, 7054–7060 (2016).

6.	Sahoo, P. K., Memaran, S., Xin, Y., Balicas, L. & Gutiérrez, H. R. One-pot growth of two-dimensional lateral heterostructures via sequential edge-epitaxy. *Nature* **553**, 63–67 (2018).

7.	Huang, Y. *et al.* Universal mechanical exfoliation of large-area 2D crystals. *Nat Commun* **11**, 2453 (2020).

8.	Liu, Y. *et al.* Promises and prospects of two-dimensional transistors. *Nature* **591**, 43–53 (2021).

9.	Shree, S., Paradisanos, I., Marie, X., Robert, C. & Urbaszek, B. Guide to optical spectroscopy of layered semiconductors. *Nat Rev Phys* **3**, 39–54 (2021).

10.	Benameur, M. M. *et al.* Visibility of dichalcogenide nanolayers. *Nanotechnology* **22**, 125706 (2011).

11.	Blake, P. *et al.* Making graphene visible. *Appl. Phys. Lett.* **91**, 063124 (2007).

12.	Dong, X. *et al.* Line-Scan Hyperspectral Imaging Microscopy with Linear Unmixing for Automated Two-Dimensional Crystals Identification. *ACS Photonics* **7**, 1216–1225 (2020).

13. Dong, X. *et al.* 3D Deep Learning Enables Accurate Layer Mapping of 2D Materials. *ACS Nano* **15**, 3139–3151 (2021).

14. Akinwande, D. *et al.* Graphene and two-dimensional materials for silicon technology. *Nature* **573**, 507–518 (2019).

15. Li, S. *et al.* Nanometre-thin indium tin oxide for advanced high-performance electronics. *Nat. Mater.* **18**, 1091–1097 (2019).

16. Tak, Y.-H., Kim, K.-B., Park, H.-G., Lee, K.-H. & Lee, J.-R. Criteria for ITO (indium–tin-oxide) thin film as the bottom electrode of an organic light emitting diode. *Thin Solid Films* **411**, 12–16 (2002).

17. Way, A. *et al.* Fluorine doped tin oxide as an alternative of indium tin oxide for bottom electrode of semi-transparent organic photovoltaic devices. *AIP Advances* **9**, 085220 (2019).

18. Gaillet, M., Yan, L. & Teboul, E. Optical characterizations of complete TFT–LCD display devices by phase modulated spectroscopic ellipsometry. *Thin Solid Films* **516**, 170–174 (2007).

19. Ma, X., Gong, Y., Wu, J., Li, Y. & Chen, J. Impacts of atomistic surface roughness on electronic transport in n-type and p-type $MoS_2$ field-effect transistors. *Jpn. J. Appl. Phys.* **58**, 110905 (2019).

20. Ma, Y. & Zhi, L. Graphene-Based Transparent Conductive Films: Material Systems, Preparation and Applications. *Small Methods* **3**, 1800199 (2019).

21. Gao, L. *et al.* Face-to-face transfer of wafer-scale graphene films. *Nature* **505**, 190–194 (2014).

22. Wang, M. *et al.* Single-crystal, large-area, fold-free monolayer graphene. *Nature* **596**, 519–524 (2021).

23. Wang, L. *et al.* Ultrathin Oxide Films by Atomic Layer Deposition on Graphene. *Nano Lett.* **12**, 3706–3710 (2012).

24. Henning, A. *et al.* Aluminum Oxide at the Monolayer Limit via Oxidant-Free Plasma-Assisted Atomic Layer Deposition on GaN. *Advanced Functional Materials* **31**, 2101441 (2021).

25. Bing, D. *et al.* Optical contrast for identifying the thickness of two-dimensional materials. *Optics Communications* **406**, 128–138 (2018).

26. Chae, W. H., Cain, J. D., Hanson, E. D., Murthy, A. A. & Dravid, V. P. Substrate-induced strain and charge doping in CVD-grown monolayer $MoS_2$. *Appl. Phys. Lett.* **111**, 143106 (2017).

27. Wei, Z. *et al.* Wafer-Scale Oxygen-Doped $MoS_2$ Monolayer. *Small Methods* **n/a**, 2100091.

28. Lee, Y.-H. *et al.* Synthesis of Large-Area $MoS_2$ Atomic Layers with Chemical Vapor Deposition. *Advanced Materials* **24**, 2320–2325 (2012).

29. Najmaei, S., Liu, Z., Ajayan, P. M. & Lou, J. Thermal effects on the characteristic Raman spectrum of molybdenum disulfide ($MoS_2$) of varying thicknesses. *Appl. Phys. Lett.* **100**, 013106 (2012).

30. Klein, J. *et al.* Robust valley polarization of helium ion modified atomically thin $MoS_2$. *2D Mater.* **5**, 011007 (2017).

31. Mignuzzi, S. *et al.* Effect of disorder on Raman scattering of single-layer $MoS_2$. *Phys. Rev. B* **91**, 195411 (2015).

32. Li, Y. *et al.* Photoluminescence of monolayer $MoS_2$ on $LaAlO_3$ and $SrTiO_3$ substrates. *Nanoscale* **6**, 15248–15254 (2014).

33. Dean, C. R. *et al.* Boron nitride substrates for high-quality graphene electronics. *Nature Nanotech* **5**, 722–726 (2010).

34. Tsoi, S. *et al.* van der Waals Screening by Single-Layer Graphene and Molybdenum Disulfide. *ACS Nano* **8**, 12410–12417 (2014).

35. Liu, H., Xu, K., Zhang, X. & Ye, P. D. The integration of high-k dielectric on two-dimensional crystals by atomic layer deposition. *Appl. Phys. Lett.* **100**, 152115 (2012).

36. Li, N. *et al.* Atomic Layer Deposition of $Al_2O_3$ Directly on 2D Materials for High-Performance Electronics. *Advanced Materials Interfaces* **6**, 1802055 (2019).

37. Ferrari, A. C. & Robertson, J. Interpretation of Raman spectra of disordered and amorphous carbon. *Phys. Rev. B* **61**, 14095–14107 (2000).

38. Ferrari, A. C., Robertson, J., Ferrari, A. C. & Robertson, J. Raman spectroscopy of amorphous, nanostructured, diamond–like carbon, and nanodiamond. *Philosophical Transactions of the Royal Society of London. Series A: Mathematical, Physical and Engineering Sciences* **362**, 2477–2512 (2004).

39. Tuinstra, F. & Koenig, J. L. Raman Spectrum of Graphite. *J. Chem. Phys.* **53**, 1126–1130 (1970).

40. Matthews, M. J., Pimenta, M. A., Dresselhaus, G., Dresselhaus, M. S. & Endo, M. Origin of dispersive effects of the Raman D band in carbon materials. *Phys. Rev. B* **59**, R6585–R6588 (1999).

41. Jerng, S. K. *et al.* Nanocrystalline Graphite Growth on Sapphire by Carbon Molecular Beam Epitaxy. *J. Phys. Chem. C* **115**, 4491–4494 (2011).

42. Dasgupta, D., Demichelis, F. & Tagliaferro, A. Electrical conductivity of amorphous carbon and amorphous hydrogenated carbon. *Philosophical Magazine B* **63**, 1255–1266 (1991).

43. Ferrari, A. C. *et al.* Stress reduction and bond stability during thermal annealing of tetrahedral amorphous carbon. *Journal of Applied Physics* **85**, 7191–7197 (1999).

44. Zhao, Y. *et al.* Interface engineering with an $AlO_x$ dielectric layer enabling an ultrastable $Ta_3N_5$ photoanode for photoelectrochemical water oxidation. *Journal of Materials Chemistry A* **9**, 11285–11290 (2021).

45. Breazu, C. *et al.* Nucleobases thin films deposited on nanostructured transparent conductive electrodes for optoelectronic applications. *Sci Rep* **11**, 7551 (2021).

# Scalable transparent conductive thin films with electronically passive interfaces for direct chemical vapor deposition of 2D materials

*Theresa Grünleitner,*[a,‡] *Alex Henning,*[a,‡,*] *Michele Bissolo,*[a] *Armin Kleibert,*[b] *Carlos A.F. Vaz,*[b] *Andreas V. Stier,*[a] *Jonathan J. Finley,*[a] *and Ian D. Sharp*[a,*]

[a] Walter Schottky Institute and Physics Department, Technical University of Munich, 85748 Garching, Germany.

[b] Swiss Light Source, Paul Scherrer Institute, CH-5232 Villigen PSI, Switzerland.

[‡]These authors contributed equally to this work.

*Authors to whom correspondence should be addressed: sharp@wsi.tum.de and Alex.Henning@wsi.tum.de

**Experimental Section**

**Material Deposition:** A Quorum Technologies Q150T system featuring pulse evaporation from a carbon rod was used to evaporate a 10 nm thick amorphous carbon (a-C) film on n$^+$ Si/SiO$_2$ substrates. The carbon layer was then coated with AlO$_x$ in a hot-wall plasma-enhanced atomic layer deposition (PE-ALD) reactor (Fiji G2, Veeco CNT) in continuous flow mode at 200 °C. In the first half-cycle, the process was conducted with a base pressure of approximately 0.09 Torr. Here, ozone was used as the oxidant and trimethylaluminum TMA (electronic grade, 99.999 %, STREM Chemicals) as the precursor for $AlO_x$ and Ar (99.9999 %, Linde) as the carrier gas. The parameters were implemented so that self-limiting growth occurs, which was controlled by monitoring the film thickness *in situ* using spectroscopic ellipsometry. In the second half-cycle, we used $H_2$ (99.9999 %, Linde) as the carrier and plasma gas and lowered the pressure to 0.02 Torr. The $H_2$ plasma generation was conducted inductively using a coupled plasma source in a copper coil, which was wrapped around the sapphire tube. To ignite and sustain the plasma, a radio frequency (rf) bias at 13.64 MHz and 100 W was applied to the copper coil. The sequence for one ALD $AlO_x$ cycle was 0.08 s TMA dose, 30 s Ar purge, 10 s $H_2$ purge, 2 s $H_2$ plasma, 10 s Ar purge, 10 s $H_2$ purge, 2 s $H_2$ plasma, 10 s Ar purge. The total thickness of the $AlO_x$ film was 1 nm. Afterwards, we directly grew single crystalline triangular shaped $MoS_2$ on top of the $AlO_x$/nc-C thin film. For this process, we used a home built chemical vapor deposition (CVD) system with a one zone horizontal tube furnace (Nabertherm RS 80/500/11) with a 5 cm inner diameter quartz tube. $MoS_2$ was grown at ambient pressure (1 atm) with $MoO_3$ (99.9995 %; Alfa Aesar) and S (99.998 %; Merck) solid precursors at a target temperature of 850 °C and a flow of 100 sccm of Ar (99.9999 %, Linde) as the carrier gas. The TCF on $Si/SiO_2$ was placed next to the $MoO_3$ precursor about 17 cm inwards from the edge of the furnace. The S precursor was placed upstream of the $MoO_3$ precursor to obtain a precursor temperature of about 180 °C. The growth time was set to 5 minutes followed by a natural cool down. For the samples with $MoS_2$ on the TCF, the annealing of the evaporated carbon happened during the CVD growth process. For all other annealed samples, we used the same CVD setup and parameters without the $MoO_3$ and S precursors.

***in situ* spectroscopic ellipsometry:** The thickness of the ALD layer was monitored in real-time during the deposition using an *in situ* spectroscopic ellipsometer (M-2000, J. A. Woollam) and a sampling time of 3 s. The Xenon light source (Hamamatsu, L2174-01) with a spot of approximately 5x8 mm$^2$ passed through a fused silica quartz window (Lesker, VPZL-275DU) at an angle of 67°. The data were fitted using a general oscillator model.

**Atomic force microscopy and scanning electron microscopy:** AFM images of $MoS_2$ flakes were acquired using a Bruker Dimension Icon XR (Bruker, USA) AFM in ambient air with platinum silicide (PtSi) probes (Nanosensors). AFM measurements for roughness determination were performed using a Bruker Multimode V microscope (Billerica, MA, USA) in ambient air with NGS30 AFM probes (TipsNano) with a nominal tip radius of 8 nm, typical force constant of 40 N/m and resonance frequency of 320 kHz. All images were acquired in tapping mode with a scan rate of 0.5 kHz and 512-point sampling for image sizes of 12×12 μm$^2$ and 3×3 μm$^2$. SEM images were collected using a Zeiss NVISION 40 with a secondary electron detector.

**Synchrotron X-ray photoemission electron microscopy:** XPS maps were measured at the SIM beamline at the Swiss Light Source (SLS), Paul Scherrer Institut (PSI), by means of X-ray photoemission electron microscopy (XPEEM using a spectroscopic low energy electron microscopy (LEEM) instrument, Elmitec GmbH). During all measurements, a bias of 15 kV was applied between the sample and the objective lens. The spatially resolved X-ray photoelectron spectroscopy images were measured by collecting XPEEM image stacks as a function of photon energy for a fixed value of the energy analyzer (set to 100 eV) to be able to keep the focus settings constant. The kinetic energy of the detected electrons determines the probing depth and was chosen to analyze monolayer $MoS_2$. The images were analyzed using a Python script, the Mo 3d core level spectra were extracted from the maps using ImageJ, and the spectra were fitted using CasaXPS. Thereby, the Mo 3d – S 2s core level spectrum (Fig. 2c) was calibrated using the adventitious carbon reference peak at 284.8 eV.

**Optical characterization:** Raman and PL measurements were performed at room temperature with a 532 nm excitation laser. For all Raman and PL measurements, we used a home-built Raman setup with a Horiba iHR 550 spectrometer (entrance slit 200 µm, 2400 grooves/mm grating) and liquid nitrogen cooled Horiba Symphony II CCD detector. For the Raman and PL data of $MoS_2$, we used a 100x objective lens, whereas for the Raman data for the different stages of sample fabrication we used a 20x objective lens. All Raman measurements were performed using an excitation power of 5.1 mW, while PL measurements of $MoS_2$ on $Si/SiO_2$ and of the TCF were collected at 28 µW.

Energy calibration of Raman spectra was carried out using the 520 $cm^{-1}$ Si Raman line. Raman spectra of $MoS_2$ were fit using pseudo-Voigt functions after averaging over three spectra containing an average over 60 sweeps each. The spectra of the nc-C/$AlO_x$ films were fit using the Breit-Wigner-Fano (BWF) function for all G peaks[37] and a Lorentzian lineshape for the D peaks. Here, the G peak position was calculated from the fitted peak position, $\omega_0$, and is smaller than the measured value because $\omega_0$ is due to the undamped mode.[37]

**Electrical characterization:** For the determination of the sheet resistivity, we defined contacts via shadow masks, electrically contacted those with probes mounted on micromanipulators, and measured the voltage drop between the inner contacts while applying a constant current at the outer contacts using a Keithley 2400 source meter unit. For each given resistivity, we averaged over four measurements.

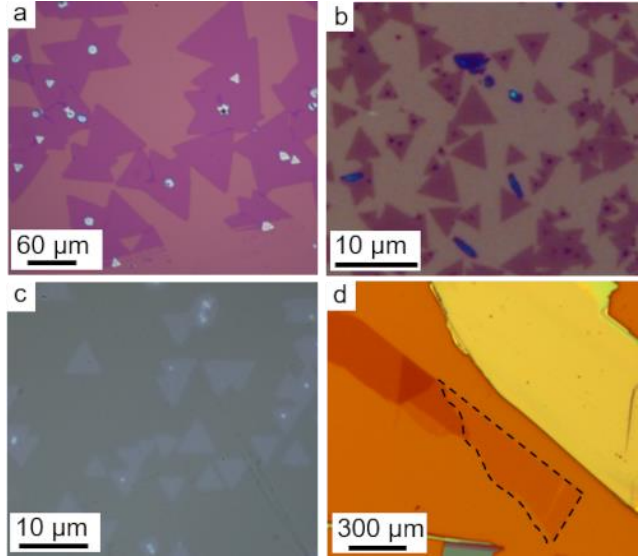

**Figure S1.** Optical microscopy images of 2D MoS$_2$ on different substrates, used to extract the optical contrast. CVD-grown MoS$_2$ on (a) Si/SiO$_2$, (b) the TCF on Si/SiO$_2$, (c) on sapphire, and (d) exfoliated MoS$_2$ on Au. The contrast of MoS$_2$ on sapphire and Au result in $C_{\text{sapphire}} = 0.08$ and $C_{\text{Au}} = 0.17$, respectively. All samples were measured using the same setup and settings, such that setup-related differences in the optical contrast are avoided.

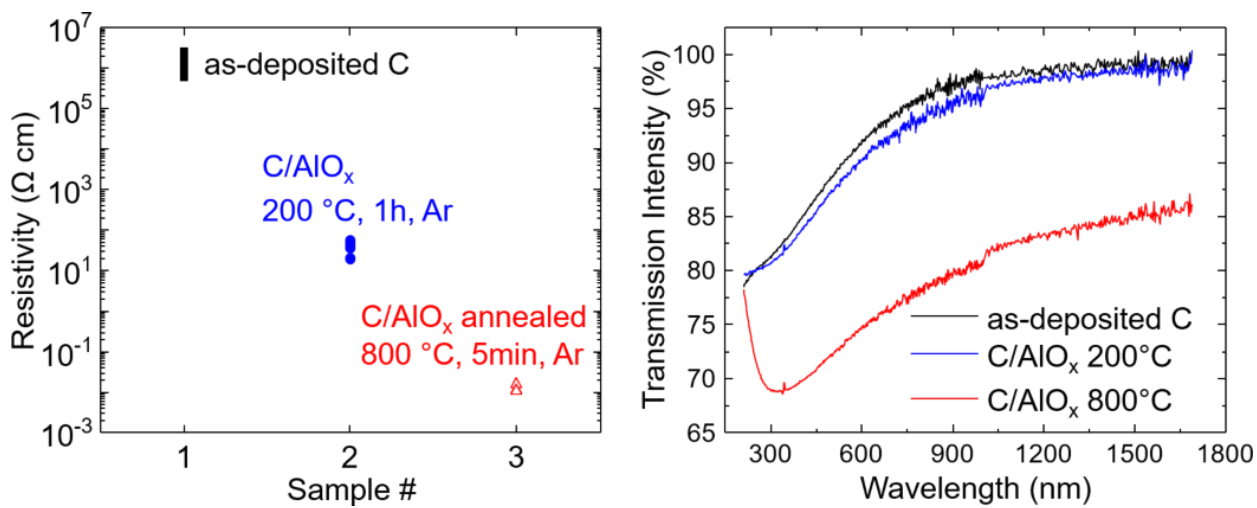

**Figure S2.** Resistivity (left) and transmission spectra (right) of the novel TCF deposited on sapphire. The transmission spectrum of sapphire with evaporated carbon (C) is shown in black, sapphire with evaporated C and ALD $AlO_x$ in blue, and the complete TCF on sapphire after annealing at 800 °C in red.

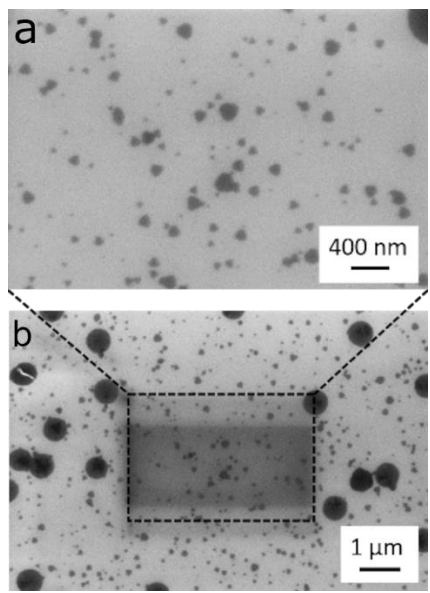

**Figure S3.** SEM images of $MoS_2$ on insulating Si/$SiO_2$ at (a) high magnification and (b) a subsequent zoom-out to examine charging effects. Poor resolution in the high magnification image is a consequence of substrate charging. In contrast, Fig. 1d shows high-resolution images of $MoS_2$

on the TCF. This comparison demonstrates a significant advantage of the TCF over the standard Si/SiO$_2$ substrate for electron-based imaging and spectroscopy.

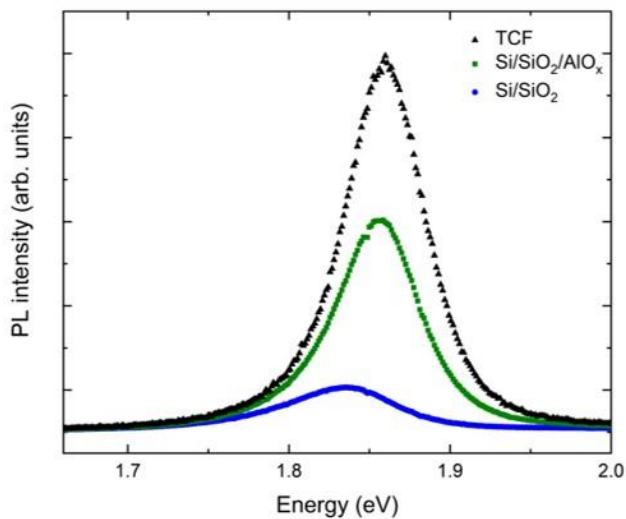

**Figure S4.** PL amplitude comparison of MoS$_2$ on Si/SiO$_2$ (blue circles), Si/SiO$_2$/AlO$_x$ (green squares), and the TCF on Si/SiO$_2$ (black triangles).

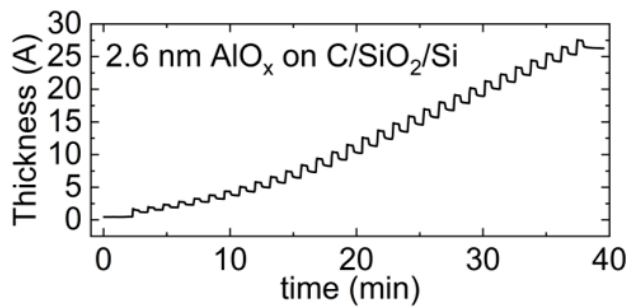

**Figure S5.** *In situ* spectroscopic ellipsometry of the atomic layer deposition of AlO$_x$ on an amorphous carbon surface.